*Article*

# Prediction of short and long-term droughts using artificial neural networks and hydro-meteorological variables

**Yousef Hassanzadeh[1], Mohammadvaghef Ghazvinian[2*], Amin Abdi[2], Saman Baharvand[3], Ali Jozaghi[3]**

[1] Department of Water Engineering, Center of Excellence in Hydroinformatics, Faculty of Civil Engineering, University of Tabriz, Tabriz, Iran; yhassanzadeh@tabrizu.ac.ir
[2] Department of Water Engineering, Faculty of Civil Engineering, University of Tabriz, Tabriz, Iran; aabdi@tabrizu.ac.ir
[3] Department of Civil Engineering, University of Texas at Arlington, P.O. Box 19019, Arlington, TX 76019, USA; Saman.Baharvand@uta.edu, Ali.Jozaghi@mavs.uta.edu
[*] Corresponding author: mohammadvaghef.ghazvinian@mavs.uta.edu ; Tel.: +1-817-437-6440; Current affiliation: Department of Civil Engineering, University of Texas at Arlington, P.O. Box 19019, Arlington, TX 76019, USA

**Abstract:** Drought is a natural creeping threat with numerous damaging effects in various aspects of human life. Accurate drought prediction is a promising step in helping policy makers to set drought risk management strategies. To fulfill this purpose, choosing appropriate models plays an important role in predicting approach. In this study, different models of Artificial Neural Network (ANN) are employed to predict short and long-term of droughts by using Standardized Precipitation Index (SPI) at different time scales, including 3, 6, 12, 24 and 48 months in Tabriz city, Iran. To this end, different combination of calculated SPI and time series of various hydro-meteorological variables, such as precipitation, wind velocity, relative humidity and sunshine hours for years 1992 to 2010 are used to train the ANN models. In order to compare the models performances, some well-known measures, namely RMSE, Mean Absolute Error (MAE) and Correlation Coefficient (CC) are utilized in the present study. The results illustrate that the application of all hydro-meteorological variables significantly improves the prediction of SPI at different time scales.

**1. Introduction and Background**

Different studies have been done on the impact of the water science and management issues on the future of the human and aquatic life-cycle (Jozaghi and Shamsai 2017; Jozaghi et al. 2018; Baharvand and Lashkar-Ara 2019; Mahmoudian et al. 2019; Martinez et al. 2019; Singh et al. 2019; Darabnoush Tehrani et al. 2019). However, with the changes in climate variables and their impact on water resources, the entire water cycle process should be studied as the requirement of the water budget investigation.

Drought "A sustained, extended deficiency in precipitation" (WMO 1986) is an expected meteorological event, impossible to avoid, causing spread spectrum of complex socio-economic and environmental impacts (Mishra and Desai 2006; Farokhnia et al. 2010; Farahmand and Aghakouchak 2015). Statistics reveal that droughts have affected large amounts of people around the globe: almost 35% of people who have suffered from natural disasters and 50% of mortalities. In terms of financial damages, droughts have cost up to 7% of global economy (Chen et al. 2013). In the last few decades, severe droughts have affected numerous regions in almost every continent on the world. For more information on global drought events and impacts in recent decades, see: Around the world (Mishra and Singh 2010), Africa (Batterbury and Warren 2001; OCHA 2011; Funk 2011; USAID/FEWSN 2011), Australia (Murphy and Timbal 2007; Bond et al. 2008; Aghakouchak et al. 2014), Asia (Agrawala et al. 2001; Guha-Sapir et al. 2004; OFDA/CRED 2008), Europe (European Communities 2007; Feyen and Dankers 2009), and United States (Riebsame et al. 1990; FEMA 1995; Wilhite and Hayes 1998; Ross and Lott 2003; Aghakouchak et al. 2014).

Like many other countries, droughts are normal feature of climate in Iran. Being located mainly in semi-arid and arid regions of the world (with long-term average precipitation 224-275 mm/year, less than one third of the world average), Iran has been exposed to severe droughts during last decades, causing complex environmental, social and economic damages (Keshavarz et al. 2013). According to the various studies, the recurring droughts could worsen the water resources of the most parts of Iran (Golian et al. 2015). In this way, there are many outstanding achievements in drought modeling of Iran (e.g., Abbaspour and Sabetraftar 2005; Morid et al. 2006, 2007; Karamouz et al. 2009; Shiau and Modarres 2009; Raziei et al. 2009, 2011; Dezfuli et al. 2010; Farokhnia et al. 2011; Tabari et al. 2012; Abdi et al. 2017a, b, c).

Traditionally, various models such as linear and multi linear regression models (Kumar and Panu 1997; Liu and Negron-Juarez 2001), stochastic models (Rao and Padmanabhan 1984; Chung and Salas 2000; Mishra and Desai 2005a; Han et al. 2010), Markov Chain and loglinear models (Lohani and Loganathan 1997; Lohani et al.1998; Paulo et al. 2005; Paulo and Pereira 2007; Moreira et al. 2006; Sen 1990; Steinemann 2003) have been used for drought prediction. Limitation of all these linear models in capturing nonlinearities and nonstationarities in hydrologic time series, in addition, complexity and uncertainty in input variables (in drought analysis: e.g., climate or/and drought indices) have made scientists to use better solutions.

Soft computing methods including Multi-Criteria Decision Making (MCDM) methods (Jozaghi et al. 2018), Artificial Neural Network (ANN) (Büyükşahin & Ertekin 2019, Kardan Moghaddam et al. 2019), fuzzy rule based systems (Jiang et al. 2019, Bose & Mali 2019, Vijayalaksmi & Babu 2015), kernel methods such as Support Vector Machine (SVM) (Shabani et al. 2017), and hybrid models have been increasingly used in hydrology and consequently drought prediction during last two decades. Kim and Valdes (2003) were first scientists to use hybrid wavelet-ANN model for drought prediction in Conchos

River Basin, Mexico using PDSI as input variable. Mishra and Desai (2006) compared SPI forecast results of ANN models and linear stochastic models (ARIMA) in India. Mishra et al. (2007) combining ARIMA and ANN introduced a new hybrid model to forecast SPI time series in the same region. Morid et al. (2007) applied ANN models to forecast time series of SPI and EDI indices in Iran. For this purpose, they used different combination of mentioned drought indices and climate indices including Southern Oscillation Index (SOI) and North Atlantic Oscillation (NAO). In a similar study, Bacanli et al. (2008) employed Adaptive Neuro-Fuzzy Interface System (ANFIS) to predict quantitative amounts of SPI in Central Anatolia, Turkey. Different combination of preceding monthly rainfall and SPI values were used as input variables. Cutore et al. (2009) used ANN to predict Palmer Hydrological Drought Index (PHDI) with 4 months lead-time in Italy. In a study performed by Farokhnia et al. (2010) Sea Level Pressure (SLP) and effective Sea Surface Temperature (SST) grids derived by data mining methods were used as predictors in ANFIS models to predict drought in Tehran plain, Iran. Ozger et al. (2012) with conjunction of wavelet transform and fuzzy logic modeled the values of Palmer Modified Drought Index (PMDI). Belayneh et al. (2014) compared the performance of five models including ARIMA, ANN, Support Vector Regression (SVR), WANN and WA-SVR to predict SPI in Ethiopia. The above-mentioned studies indicate the superiority of the soft computing methods compared to the traditional methods.

In the present study, time series of SPI at multiple time scales and hydro-meteorological variables including monthly precipitation, wind velocity, relative humidity, and sunshine hours for a period of 18 years from 1992 to 2009 were used as input data for ANN method, we predicted short and long-term droughts in Tabriz city, Iran.. For this purpose, different models of the ANN were analyzed and the best model was selected according to three well-known models performances namely RMSE, MAE, and CC. The rest of this paper is organized as follows. Section 2 describes study area and methods used. Section 3 presents the results. Section 4 provides the conclusions and future research recommendations.

## 2. Methodology

*2.1. Case Study*

In this study, monthly hydro-meteorological data of Tabriz synoptic station (38.05 °E and 46.17 °N) including precipitation (P), relative humidity (H), wind velocity (V) and sunshine hours (S) from years 1992 to 2009 are utilized to predict drought. The data mentioned above were obtained from the East-Azarbaijan Meteorological Organization. Tabriz city, the capital of East Azerbaijan province, Iran (Fig. 1) with the population of approximately 1.5 million is one of major industrial cities in Iran. This city is located in a semi-arid area where the mean annual precipitation and temperature amounts are 290 mm and 12.5 °C, respectively (Zarghami et al. 2011). The time series of hydro-meteorological data are presented in Fig. 2.

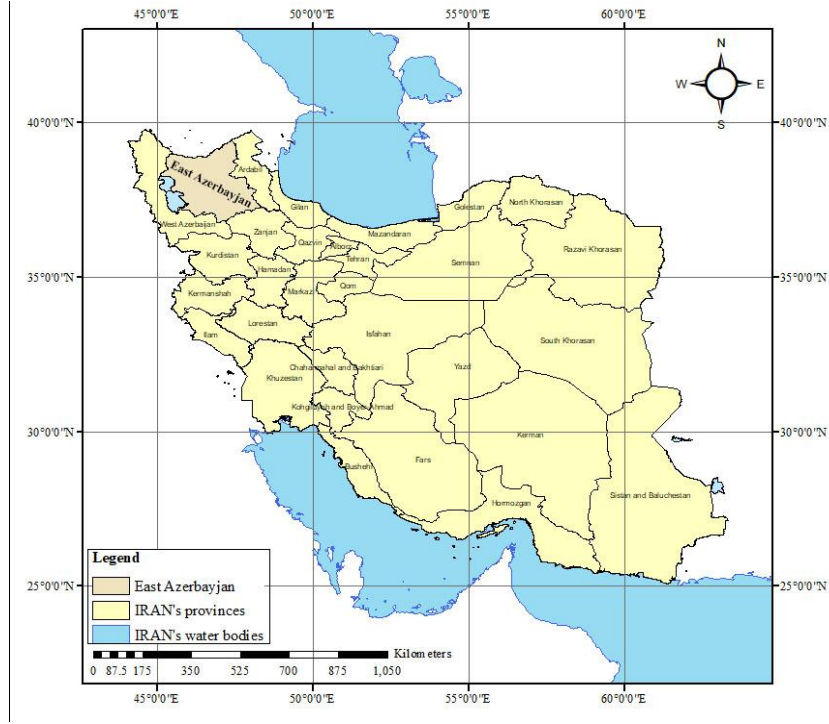

**Fig. 1.** East Azerbaijan province location on Iran's map

*2.2 Standardized Precipitation Index (SPI)*

During last decades, several indices have been developed to describe different aspects of droughts. Indices facilitate the evaluation of various characteristics of drought (Wilhite et al. 2000). Standardized precipitation index introduced by Mckee et al. (1993), has a simple structure and does not depend on location or any other meteorological data except for precipitation. It can be calculated for short and long-term at different time scales (e.g., 3, 6, 12, 24 and 48 months). SPI provides unique opportunity to assess different drought types (i.e., meteorological, hydrological, agricultural) and has a standard nature to compare drought conditions in different regions and climates (Farahmand and Aghakouchak 2015; Damberg and Aghakouchak 2013; Raziei et al. 2009; Abdi et al. 2017e). Table 1 provides SPI based drought classification and probabilities introduced by Lloyd- Hughes and Saunders (2002).

*2.2.1 SPI time series calculation based on precipitation data*

Following Mishra and Desai (2006), Gamma distribution is used to fit to the frequency distribution of precipitation summed over individual month (for various time scales). The gamma probability density function is defined as:

$$g(x) = \frac{1}{\beta^{\alpha}\Gamma(\alpha)} x^{\alpha-1} e^{\frac{-x}{\beta}} \quad \text{for } x > 0 \tag{1}$$

where $\alpha$ and $\beta > 0$ are shape and scale parameters, respectively, and $x > 0$ is the precipitation amount. The following improper integral defines the gamma function:

$$\Gamma(\alpha) = \int_0^\infty y^{\alpha-1} e^{-y} dy \tag{2}$$

In order to fit the gamma distribution to the precipitation data series, it is necessary to calculate both parameters. Thom (1958) approximation equations (eq. (3) to (6)) for parameter estimation are employed following Edwards and Mckee (1997).

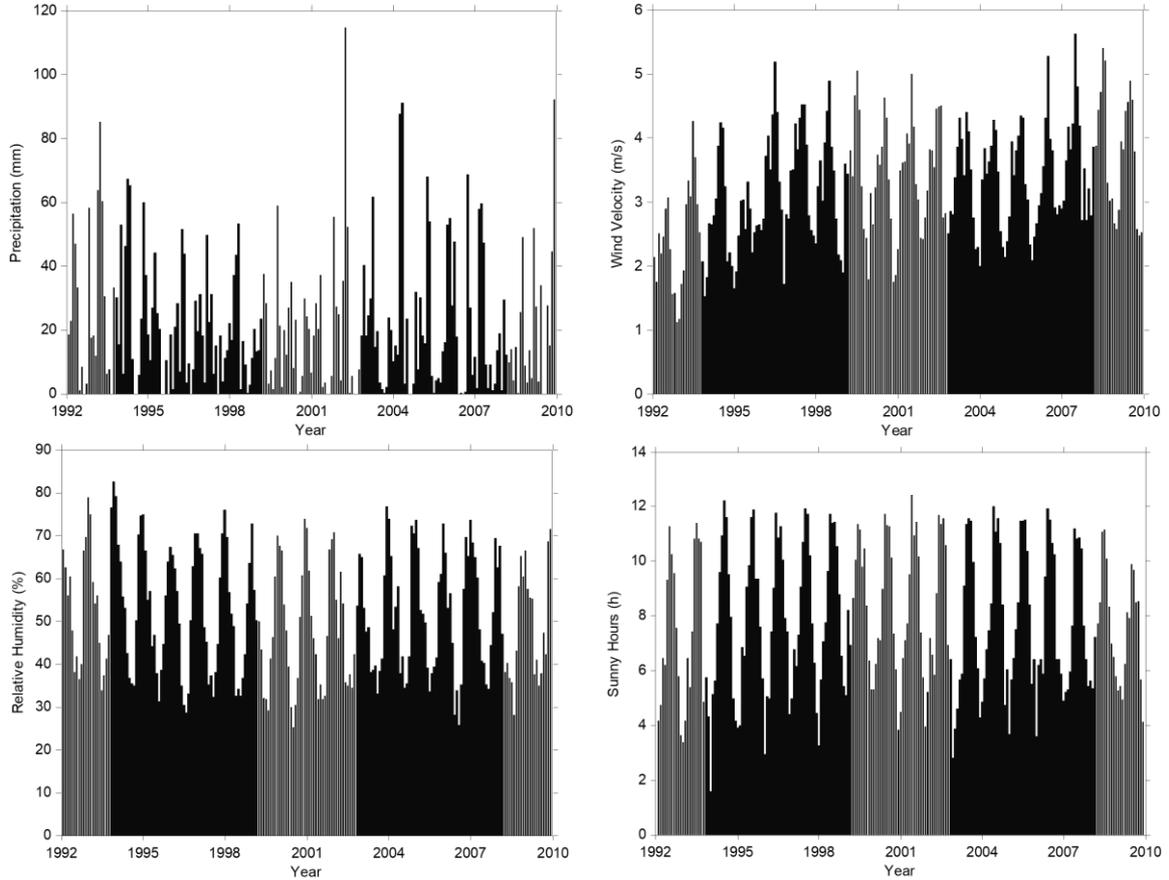

**Fig. 2** Time series of precipitation (upper left), wind velocity (upper right), relative humidity (lower left) and sunshine hours (lower right) data of Tabriz synoptic station (1992-2010)

**Table 1** Drought classification by SPI value and corresponding event probabilities (Lloyd-Hughes and Saunders, 2002)

| SPI value | Category | Probability (%) |
|---|---|---|
| 2.00 or more | Extremely wet | 2.3 |
| 1.50 to 1.99 | Severely wet | 4.4 |
| 1.00 to 1.49 | Moderately wet | 9.2 |
| 0 to 0.99 | Mildly wet | 34.1 |
| 0 to −0.99 | Mild drought | 34.1 |
| −1.00 to −1.49 | Moderate drought | 9.2 |
| −1.50 to −1.99 | Severe drought | 4.4 |
| −2 or less | Extreme drought | 2.3 |

$$\hat{\alpha} = \frac{1}{4A}\left(1+\sqrt{1+\frac{4A}{3}}\right) \quad (3)$$

$$\hat{\beta} = \frac{\bar{x}}{\hat{\alpha}} \quad (4)$$

$$A = \ln(\bar{x}) - \frac{\sum \ln(x)}{n} \quad (5)$$

where $\bar{x}$ is the mean of precipitation distribution and n is the number of precipitation records. Gamma cumulative probability G(x) of a desired precipitation value, now can be calculated as following equation (Mishra and Desai 2006; Lloyd-Hughes and Saunders 2002).

$$G(x) = \int_0^x g(x)dx = \frac{1}{\hat{\beta}^{\hat{\alpha}}\Gamma(\hat{\alpha})}\int_0^x x^{\hat{\alpha}} e^{\frac{-x}{\hat{\beta}}}dx \quad (6)$$

Since the observed precipitation time series may contain zeros, according to Mishra and Desai (2006) the cumulative probability, H(x), can be expressed by:

$$H(x) = q + (1-q)G(x) \quad (7)$$

$$q = m/n \quad (8)$$

In the above equation, m and n are the number of zero precipitation and sample size, respectively. The cumulative probability is then transformed to standard normal variable (Z) with mean zero and variance one (Lloyd-Hughes and Saunders 2002). For this purpose, graphical and analytical methods (Lloyd-Hughes and Saunders 2002) have been suggested. In this study the analytical approximation of Abramowitz and Stegun (1965), which is routinely used to derive Z values, has been employed. Related equations can be found in Mishra and Desai (2006) and Lloyd-Hughes and Saunders (2002).

*2.3 Artificial Neural Networks (ANN)*

Artificial neural networks helps to understand the brain, human cognition, and perception and have been proven to be successful for solving different problems in pattern classification, decision making, and predicting (Fine 1999). Among various ANN models and training algorithms, multi-layer perceptron neural network model (MLP) trained with feed-forward backpropagation algorithm have been extensively used for simulating and predicting purposes in hydrology and water resources (ASCE 2000a, b; Maier and Dandy 2000). MLPs consist of an input layer (data are presented to the network via input layer), hidden layer(s) (where the actual processing is done via a system of weighted connections) and an output layer (where the answer is produced). Individual neurons in a layer are connected with different weights to every neuron in the following layer (Belayneh et al. 2016). Weighted inputs are mapped to the output of individual neurons by a linear or nonlinear activation function (Fig. 2).

According to Kim and Valdes (2003) the output value of the above network is given by:

$$\hat{y}_k = f_o\left[\sum_{j=1}^{m} w_{kj} \cdot f_h\left(\sum_{i=1}^{n} w_{ji} x_i + w_{jo}\right) + w_{ko}\right] \quad (9)$$

where n and m are the number of samples and neurons in hidden layer, respectively. In addition, xi is the i-th input data, $w_{ji}$ is the weight connecting the i-th neuron in the input layer and the j-th neuron in the hidden layer, $w_{jo}$ is bias for the j-th neuron in the hidden layer, $f_h$ is the activation function used in the hidden layer, $w_{kj}$ is the weight connecting the j-th neuron in the hidden layer and kth neuron in the output layer. $w_{ko}$ is bias for the kth neuron in the output layer, and $f_o$ is activation function for the output layer (Kim and Valdes 2003).

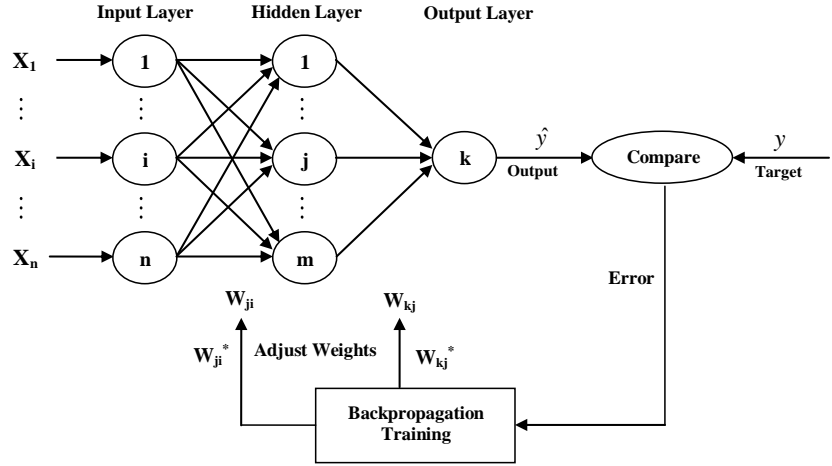

**Fig. 2** Typical three layer feed-forward neural networks with backpropagation training algorithm (Zarghami et al. 2011)

As shown in Fig .3, the error value is calculated based on the output and real values and then back propagates to the network to adjust connection weights using training algorithms. The objective of using training algorithms is to optimize the parameters of output function (weights in Eq. 9) so that the E(x,w) value i.e. network's global errors, becomes minimal.

$$E(x,w) = \sum_{p=1}^{P} E_p \quad (10)$$

$$E(p) = \frac{1}{2}\sum_{k=1}^{K}(y_k - \hat{y}_k)^2 \quad (11)$$

where P is the total number of training patterns, K is the total number of output neurons (in this study equals 1), $y_k$ is desired output at kth neuron and $\hat{y}_k$ is actual output at the kth neuron (Kim and Valdes 2003; Kisi and Uncuoglu 2005). For this purpose, second order Levenberg-Marquardt (LM) (Levenberg 1944; Marquardt 1963) back propagation algorithm is used for neural networks training. By combination of gradient descent method (also known as back propagation method) and the Gauss–Newton method, the LM algorithm solves the problems existing in both methods for neural networks training. The stability, fast convergence and less easily falling into the local minima trap has made the LM algorithm one of the most efficient training algorithms (Hagan and Menhaj 1994; Sun et al 2016).

In this research, feed-forward ANN models comprised of one input layer, one hidden layer and one output layer are used to predict SPI values. All models are trained using LM backpropagation algorithm. The activation functions for both hidden and output layers are tangent sigmoid (tansig) function. The numbers of hidden layer's neurons are selected using a trial-and-error approach. ANN models are developed using MATLAB (R2014a) ANN toolbox. For all models, first 75% of input data time series are used for training and the next 25% for validation. To this end, all input data are normalized using following equations:

$$\begin{cases} Y_i = \dfrac{X_{oi}}{X_{o\max}} \; , \; X_{oi} \geq 0 \\ Y_i = \dfrac{X_{oi}}{|X_{o\min}|} \; , \; X_{oi} < 0 \end{cases} \quad (12)$$

Where $X_{oi}$ is the observed value, $X_{o\,min}$ and $X_{o\,max}$ are respectively the minimum and maximum data in input time series (Zarghami et al., 2011).

*2.4 Model performance measures*

In this study, to evaluate the performances of all ANN models, following goodness of fit measures namely Correlation Coefficient (CC), Root Mean Square Error (RMSE) and Mean Absolute Error (MAE) are used (Wang et al. 2009; Hassanzadeh et al. 2011; Abdi et al. 2017d; Jozaghi et al. 2019).

$$CC = \dfrac{\sum_{i=1}^{n}\left[(X_{oi} - \bar{X}_o)(X_{ci} - \bar{X}_c)\right]}{\sqrt{\left[\sum_{i=1}^{n}(X_{oi} - \bar{X}_o)^2 \sum_{i=1}^{n}(X_{ci} - \bar{X}_c)^2\right]}} \quad (13)$$

$$RMSE = \sqrt{\left[\dfrac{1}{n}\sum_{i=1}^{n}(X_{oi} - X_{ci})^2\right]} \quad (14)$$

$$MAE = \dfrac{1}{n}\sum_{i=1}^{n}|X_{oi} - X_{ci}| \quad (15)$$

Where $X_{oi}$ and $X_{ci}$ are observed and predicted SPI values, n is the number of samples, $\bar{X}_0$ and $\bar{X}_c$ are the mean values taken over n.

**3. Results and discussion**

In this research various ANN models with different combination of SPI and hydro-meteorological variables and different architectures (different number of neurons in hidden layer) are considered for predicting short and long-term drought. For this purpose, at first, the values of SPI at different time scales (i.e., 3, 6, 12, 24 and 48 months) are calculated based on observed precipitation data of Tabriz synoptic station. Calculated SPI values are shown in Fig. 3. In addition, the probability of drought occurrences related to drought category are computed by using the SPI values at different time scales (Table 2).In this study, by using more than two preceding months of hydro-meteorological parameters, the models lead to more complexity due to increasing the input neurons and cause overfitting in training phase. So all models are based on two preceding month values of predictors. To highlight the study results, among all models, models with better performances (qualified models) are selected to be compared with each other. The aim of this process was two fold: first, to investigate the significance of individual predictors in SPI predicting. Second, to examine the models performance to predict the SPI with different time scales.

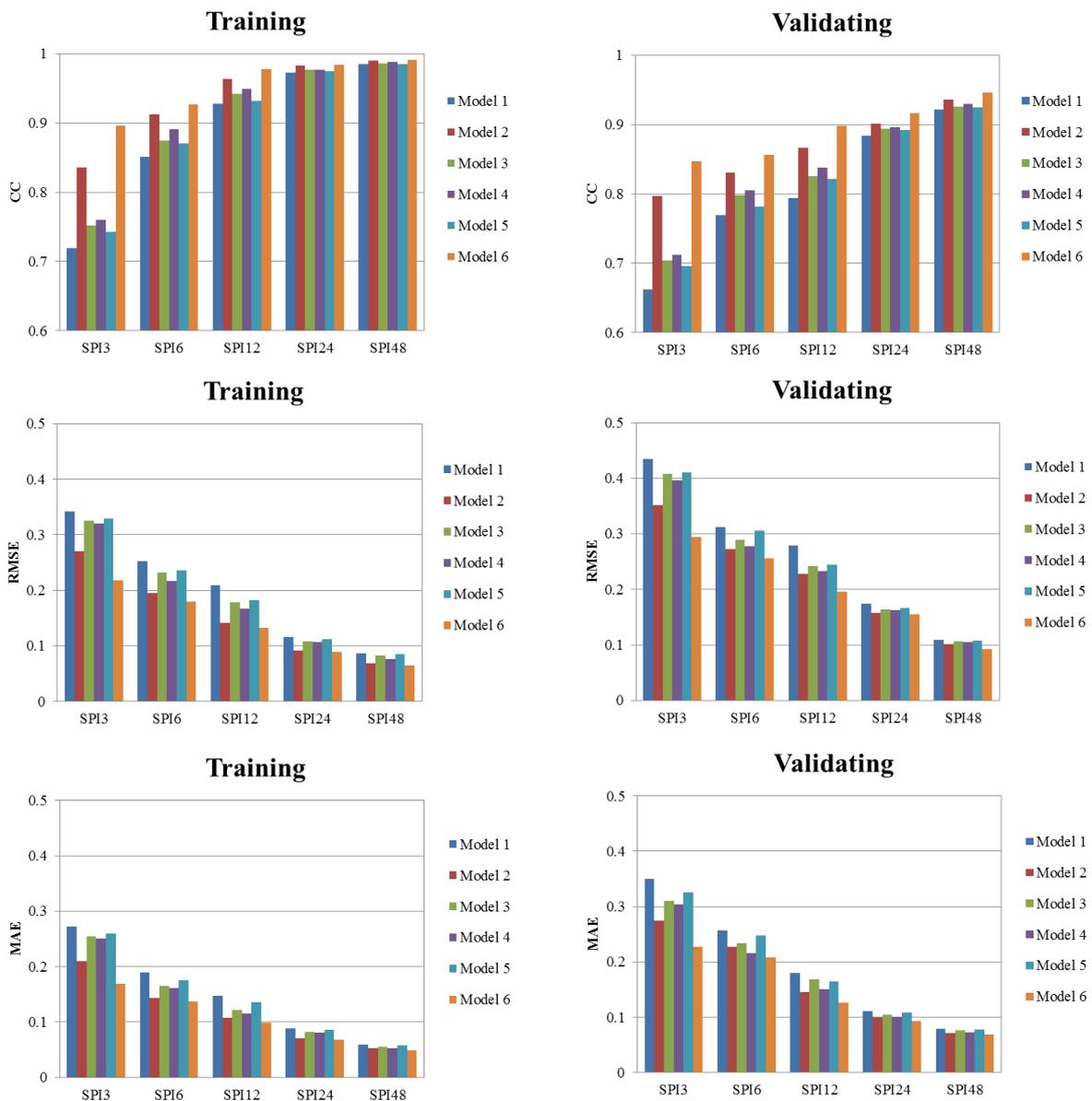

**Fig. 4** Model performance measures for SPI prediction at different time scales

**Table 2** Probability of drought occurrences related to drought category

| Drought category | Probability of drought occurrences (%) | | | | |
|---|---|---|---|---|---|
| | SPI3 | SPI6 | SPI12 | SPI24 | SPI48 |
| Mild | 43.5 | 39.8 | 48.6 | 48.1 | 39.4 |
| Moderate | 9.7 | 15.3 | 15.7 | 29.2 | 27.3 |
| Severe | 2.8 | 4.6 | 3.7 | 5.6 | 17.6 |
| Extreme | 2.3 | 1.4 | 2.3 | 0.9 | 0.0 |

*3.1 Model selection results*

Among various tested models with only one type of predictors (i.e. SPI, V, H, P, S), model 1 comprised of SPI values of two preceding months showed the best performance in predicting SPI in different time scales both in training and validation phases. This test illustrates the significance of SPI itself in precise predicting. In addition, different models with different combination of two types of predictors were examined. An analogy between all model performance criteria indicated that the presence of SPI index in models significantly improves the forecast results. Models numbered 2 to 5 with better performances in this category are illustrated in Table 3. Finally model 6 with combination of all 5 predictors is presented to be compared with other models. Qualified models are illustrated in Table 3 where P is monthly average precipitation amount (mm), H is relative humidity (%), V is the monthly average wind velocity (m/sec), S is the monthly average sunshine hours (hours).

The values of the model performance measures for SPI prediction at different time scales are presented in Fig. 4. Among six presented models in Table 3, model 6 provided the most accurate predictions of all time scales in terms of model performance measures. Models 2, 4, 3, 5, respectively showed better performances and model 1 which was based on the drought index alone had the lowest performance of all. Model ranking results based on their performances emphasize the significance of using hydro-meteorological variables in SPI prediction. The forecast results indicate that, models 2 and 4 which employ precipitation and relative humidity in addition to SPI itself, performed better than model 3 and 5 which employ average wind velocity and sunshine hours. This analogy clarifies the importance of precipitation and humidity in SPI predicting than two other predictors. A holistic review also indicates that the precipitation and sunshine hours, respectively have the most and least impact on forecast results improvement. In terms of performance measures, as shown in Fig. 4 all predicting models performed better in SPI predicting with larger time scales. Results illustrate that the increase in time scales of predicted SPI will increase the correlation coefficient (CC) values of all models and decrease RMSE and MAE values in both training and validating. According to Fig. 4, a significant convergence in individual performance measures values of all models are detected in larger time scales. In other words, by increasing the time scale of SPI index, the monthly fluctuations of the time series are decreased. So, the performance criteria of the considered models are increased for larger SPI scale.

Fig. 5 displays the observed versus predicted SPI at different time scales for both training and validation phases in best model (model 6). As it can be seen from this figure, changes of the SPI values over the times have an important effect on the accuracy of the training data. So, the efficiency of the model 6 is improved with increasing the time scales of SPI from 3 to 48 months. In addition, the frequencies of the predicted drought based on the best model are presented in Table 4. Results of this table indicate that the model 6 provides high accuracy in identification of drought probabilities.

**Table 3** Qualified ANN models with the best architecture considered in this study

| Model No. | ANN Model | Architecture |
|---|---|---|
| 1 | $SPI_t = f(SPI_{t-1}, SPI_{t-2})$ | 2-5-1 |
| 2 | $SPI_t = f(SPI_{t-1}, SPI_{t-2}, P_{t-1}, P_{t-2})$ | 4-4-1 |
| 3 | $SPI_t = f(SPI_{t-1}, SPI_{t-2}, V_{t-1}, V_{t-2})$ | 4-4-1 |
| 4 | $SPI_t = f(SPI_{t-1}, SPI_{t-2}, H_{t-1}, H_{t-2})$ | 4-4-1 |
| 5 | $SPI_t = f(SPI_{t-1}, SPI_{t-2}, S_{t-1}, S_{t-2})$ | 4-4-1 |
| 6 | $SPI_t = f(SPI_{t-1}, SPI_{t-2}, P_{t-1}, P_{t-2}, V_{t-1}, V_{t-2}, H_{t-1}, H_{t-2}, S_{t-1}, S_{t-2})$ | 10-3-1 |

**Table 4** Predicted drought frequency based on Model 6

| Drought category | Probability of drought occurrences (%) | | | | |
|---|---|---|---|---|---|
| | SPI3 | SPI6 | SPI12 | SPI24 | SPI48 |
| Mild | 39.4 | 36.0 | 43.8 | 45.9 | 38.6 |
| Moderate | 12.6 | 17.8 | 20.1 | 30.8 | 26.2 |
| Severe | 5.1 | 5.1 | 4.7 | 7.9 | 18.7 |
| Extreme | 3.2 | 1.4 | 3.1 | 0.5 | 0.0 |

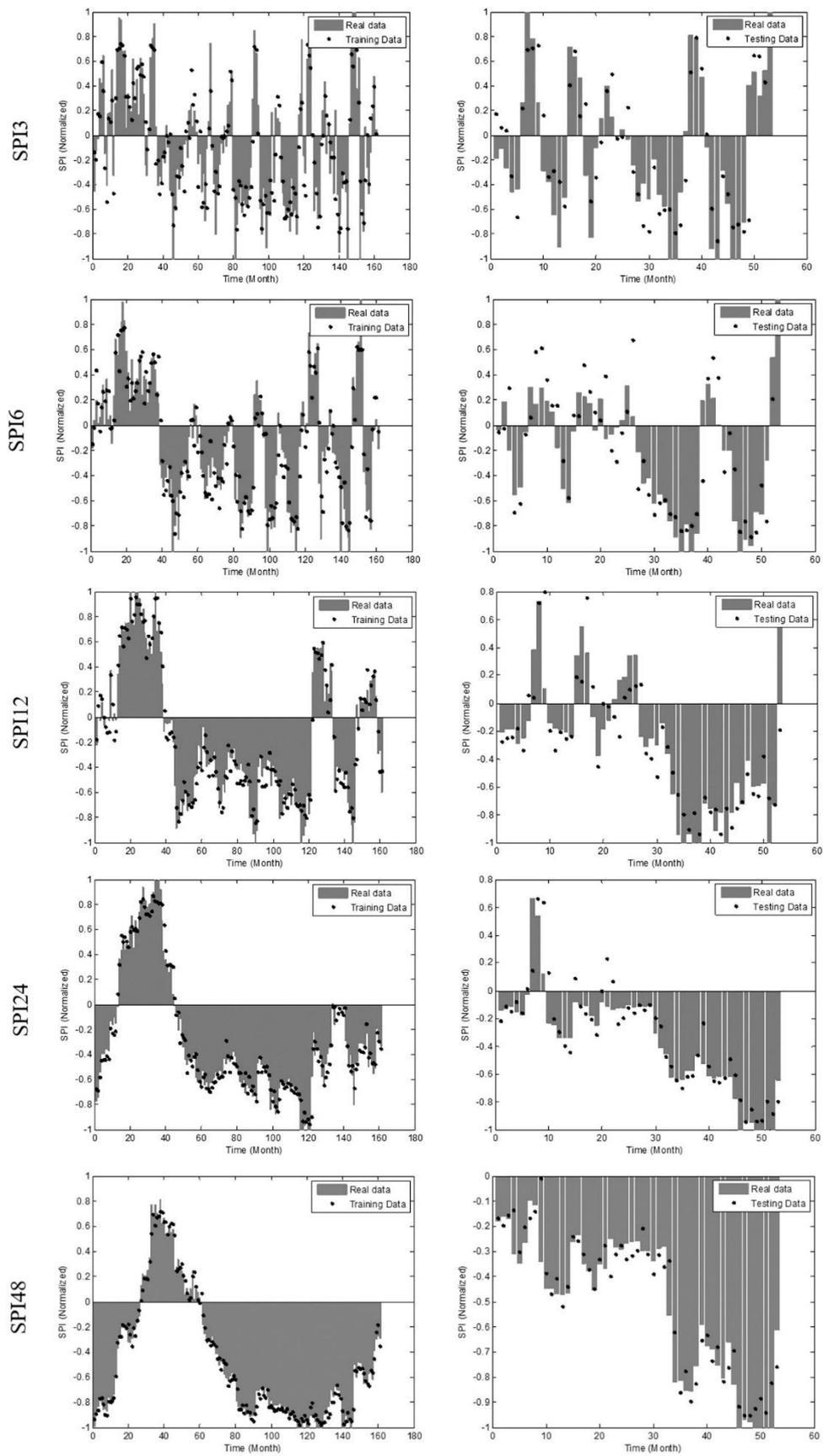

**Fig. 5** Observed versus predicted SPI at different time scales in training and validating phases of model 6

## 4. Conclusion

In this study, artificial neural networks were used to predict short and long-term drought in Tabriz, Iran. To this end, in first step, SPI at different time scales were calculated based on the precipitation time series. In addition to SPI, various hydro-meteorological variables including precipitation, wind velocity, sunshine hours and humidity for a period of 18 years from 1992 to 2010 were utilized in predicting approach. The SPI time series were predicted using two preceding month values of above-mentioned variables. In this study, various models were developed based on different combination of inputs and different architectures. In order to evaluate the performance of each model, three measures namely CC, RMSE and MAE were utilized. Forecast results of all models illustrated the significance of SPI itself in accurate prediction and the role of hydro-meteorological variables in drought prediction improvement. Model performances results indicated that precipitation data and humidity were superior to other used variables in predicting and significantly improved the performance of drought prediction. Prediction Statistics of qualified models revealed that the model using all mentioned input variables simultaneously, leads to the lowest RMSE and MAE and highest CC among all models. Accurate drought prediction can facilitate the mitigation of its devastating impacts by helping governments to better manage water resources systems. Although artificial neural networks have proved to be successful in drought prediction, having access to long term, up to date and reliable meteorological data is inevitable for more accurate predictions.

Recommendations for future research:

- Comparing developed ANN models with time series modeling approaches
- Considering spatiotemporal variables may lead to more accurate results
- Comparing the results of implementing adaptive network-based fuzzy inference system (ANFIS) with ANN

**Funding**: This research received no external funding.

**Conflicts of Interest**: The authors declare no conflict of interest.

**Acknowledgements**: The authors are grateful to East-Azarbaijan Meteorological Organization for providing the hydro-meteorological data.

## References

Abbaspour M, Sabetraftar A (2005) Review of cycles and indices of drought and their effect on water resources, ecological, biological, agricultural, social and economical issues in Iran. Int J Environ Stud 62(6):709-724. DOI: 10.1080/00207230500288968

Abdi A, Hassanzadeh Y, Ouarda TBMJ (2017a) Regional frequency analysis using growing neural gas network. J Hydrol 550:92-102. DOI: 10.1016/j.jhydrol.2017.04.047

Abdi A, Hassanzadeh Y, Talatahari S, Fakheri-Fard A, Mirabbasi R (2017b) Multivariate regional frequency analysis: Two new methods to increase the accuracy of measures. Adv Water Resour 107:290-300. DOI: 10.1016/j.advwatres.2017.07.006

Abdi A, Hassanzadeh Y, Talatahari S, Fakheri-Fard A, Mirabbasi R (2017c) Parameter estimation of copula functions using an optimization-based method. Theor Appl Climatol 129(1-2):21-32. DOI: 10.1007/s00704-016-1757-2.

Abdi A, Hassanzadeh Y, Talatahari S, Fakheri-Fard A, Mirabbasi R (2017d) Regional bivariate modeling of droughts using L-comoments and copulas. Stoch Environ Res Risk Assess 31(5):1199-1210. DOI: 10.1007/s00477-016-1222-x.

Abdi A, Hassanzadeh Y, Talatahari S, Fakheri-Fard A, Mirabbasi R (2017e) Regional drought frequency analysis using L-moments and adjusted charged system search. J Hydroinform 19(3):426-442. DOI: 10.2166/hydro.2016.228

Abramowitz M, Stegun A (1965) Handbook of mathematical formulas, graphs, and mathematical tables. Dover Publications, Inc, New York

AghaKouchak A, Feldman D, Stewardson MJ, Saphores JD, Grant S, Sanders B (2014) Australia's Drought: Lessons for California. Science 343(6178):1430-1431

Agrawala S, Barlow M, Cullen H, Lyon B (2001) The drought and humanitarian crisis in Central and Southwest Asia: a climate perspective, IRI special report N. 01-11. International Research Institute for Climate Prediction, Palisades, p 24

ASCE Task Committee on Application of Artificial Neural Networks in Hydrology (2000a) Artificial neural networks in hydrology. I. Preliminary concepts. J Hydrol Eng 5(2):115-123. DOI: 10.1061/(ASCE)1084-0699(2000)5:2(115)

ASCE Task Committee on Application of Artificial Neural Networks in Hydrology (2000b) Artificial neural networks in hydrology. II. Hydrologic applications. J Hydrol Eng 5(2): 124-137. DOI: 10.1061/(ASCE)1084-0699(2000)5:2(124)

Bacanli UG, Firat M, Dikbas EF (2009) Adaptive Neuro-Fuzzy Inference System for drought forecasting. Stoch Environ Res Risk Assess 23(8):1143–1154. DOI: 10.1007/s00477-008-0288-5

Baharvand, S., and Lashkar-ara, B. (2019). Determining the Effective of Resting Pool Area in Vertical Slot Fishways Type 1 to Pass Chinook Salmon. Journal of Civil and Environmental Engineering, 48(93), 1–12.

Batterbury SPJ, Warren A (2001) The African Sahel 25 years after the great drought: assessing progress and moving towards new agendas and approaches. Global Environ Chang 11(1):1-8. DOI: 10.1016/S0959-3780(00)00040-6

Belayneh A, Adamowski J, Khalil B, Ozga-Zielinski B (2014) Long term SPI drought forecasting in the Awash River Basin in Ethiopia using wavelet-support vector regression models. J Hydrol 508:418–429. DOI: 10.1016/j.jhydrol.2013.10.052

Belayneh A, Adamowski J, Khalil B (2016) Short-term SPI drought forecasting in the Awash River Basin in Ethiopia using wavelet transforms and machine learning methods. Sustain Water Resour Manag 2:87-101. DOI: 10.1007/s40899-015-0040-5

Bond NR, Lake PS, Arthington AH (2008) The impacts of drought on freshwater ecosystems: an Australian Perspective. Hydrobilogia 600:3–16. DOI: 10.1007/s10750-008-9326-z

Bose, M., & Mali, K. (2019). Designing fuzzy time series forecasting models: A survey. International Journal of Approximate Reasoning. https://doi.org/10.1016/j.ijar.2019.05.002

Büyükşahin, Ü. Ç, Ertekin, S. (2019). Improving forecasting accuracy of time series data using a new ARIMA-ANN hybrid method and empirical mode decomposition, Neurocomputing, https://doi.org/10.1016/j.neucom.2019.05.099.

Chen Lu, Singh VP, Guo S, Mishra AK, Guo J (2013) Drought analysis using Copulas. J Hydrol Eng 18(7):797–808. DOI: 10.1061/(ASCE)HE.1943-5584.0000697

Chung CH, Salas JD (2000) Return period and risk of droughts for dependent hydrologic processes. J Hydrol Eng 5 (3):259–268

Cutore P, Di Mauro G, Cancelliere A (2009) Forecasting palmer index using neural networks and climatic indexes. J Hydrol Eng 14:588–595. DOI: 10.1061/ASCEHE.1943-5584.0000028

Damberg L, AghaKouchak A (2014) Global trends and patterns of droughts from space. Theor Appl Climatol 117:441–448. DOI: 10.1007/s00704-013-1019-5

Darabnoush Tehrani, Amin & Kohankar Kouchesfehani, Zahra & Najafi, Mohammad & Syar, Jeffrey & Kampbell, Ed. (2019). Evaluation of Filling the Valleys of Corrugated Metal Pipes by Trenchless Spray Applied Pipe Linings. 10.1061/9780784482506.009.

Dezfuli AK, Karamouz M, Araghinejad S (2010) On the relationship of regional meteorological drought with SOI and NAO over southwest Iran. Theor Appl Climatol 100:57–66. DOI: 10.1007/s00704-009-0157-2

Edwards DC, McKee TB (1997) Characteristics of 20th century droughts in the United States at multiple time scales. Climatology Report, 97–2, Department of Atmospheric Sciences. Colorado State University, Fort Collins, p 155

European Communities (2007) Addressing the challenge of water scarcity and droughts in the European Union Commun Com 414 Final, Brussels

Farahmand A, AghaKouchak A (2015) A Generalized framework for deriving nonparametric standardized drought indicators. Adv Water Resour 76:140-145. DOI: 10.1016/j.advwatres.2014.11.012

Farokhnia A, Morid S, Byun HR (2010) Application of global SST and SLP data for drought forecasting on Tehran plain using data mining and ANFIS techniques. Theor Appl Climatol 104 (1):71-81. DOI 10.1007/s00704-010-0317-4

FEMA (1995) National mitigation strategy: Partnerships for building safer communities. Federal Emergency Management Agency, Washington DC

Fine TL (1999) Feedforward Neural Network Methodology. Springer-Verlag, New York

Feyen L, Dankers R (2009) Impact of global warming on streamflow drought in Europe. J Geophys Res 114: D17. DOI: 10.1029/2008JD011438

Funk C (2011) We thought trouble was coming. Nature 476(7358):7

Golian S, Mazdiyasni O, AghaKouchak A (2015) Trends in meteorological and agricultural droughts in Iran. Theor Appl Climatol 119:679-688. DOI: 10.1007/s00704-014-1139-6

Guha-Sapir D, Hargitt D, Hoyois P (2004) Thirty years of natural disasters 1974–2003: the numbers. Univ. Louvain Presses, Louvain

Hagan MT, Menhaj M (1994) Training feedforward networks with the Marquardt algorithm. IEEE T Neural Networ 5(6): 989–993. DOI: 10.1109/72.329697

Han P, Wang PX, Zhan SY, Zhu DH (2010) Drought forecasting based on the remote sensing data using ARIMA models. Math Comput Model 51(11–12):1398–1403. DOI: 10.1016/j.mcm.2009.10.031

Hassanzadeh Y, Abdi A, Talatahari S, Singh VP (2011) Metaheuristic algorithms for hydrologic frequency analysis. Water Resour Manag. 25(7):1855–1879. DOI: 10.1007/s11269-011-9778-1

Jiang, Z., Wu, W., Qin, H., Hu, D., & Zhang, H. (2019). Optimization of fuzzy membership function of runoff forecasting error based on the optimal closeness. Journal of Hydrology. https://doi.org/10.1016/j.jhydrol.2019.01.009

Jozaghi, A., Alizadeh, B., Hatami, M., Flood, I., Khorrami, M., Khodaei, N., & Ghasemi Tousi, E. (2018). A Comparative Study of the AHP and TOPSIS Techniques for Dam Site Selection Using GIS: A Case Study of Sistan and Baluchestan Province, Iran. Geosciences. https://doi.org/10.3390/geosciences8120494

Jozaghi, A., Nabatian, M., Noh, S., Seo, D. J., Tang, L., Zhang, J. (2019). Improving Multisensor Precipitation Estimation via Adaptive Conditional Bias-Penalized Merging of Rain Gauge

Data and Remotely-Sensed Quantitative Precipitation Estimates. submitted to the Journal of Hydrometeorology.

Jozaghi, A., Shamsai, A. (2017). Application of Geospatial Information System and Technique for Order Preference by Similarity to Ideal Solution for sitting water reservoirs Case study: South of Sistan&Balouchestan Province. Scientific- Research Quarterly of Geographical Data (SEPEHR), 25(100), pp. 5-15.(in Persian). doi: 10.22131/sepehr.2017.24802

Karamouz M, Rasouli K, Nazil S (2009) Development of a hybrid index for drought prediction: case study. J Hydrol Eng 14:617–627. DOI: 10.1061/ASCEHE.1943-5584.0000022

Kardan Moghaddam, H., Kardan Moghaddam, H., Rahimzadeh Kivi, Z., Bahreinimotlagh, M., Alizadeh,M. J. (2019). Developing comparative mathematic models, BN and ANN for forecasting of groundwater levels, Groundwater for Sustainable Development, https://doi.org/10.1016/j.gsd.2019.100237.

Keshavarz M, Karami E, Vanclay F (2013) The social experience of drought in rural Iran. J Land Use Policy 30:120–129. DOI: 10.1016/j.landusepol.2012.03.003

Kim T, Valdes JB (2003) Nonlinear model for drought forecasting based on a conjunction of wavelet transforms and neural networks. J Hydrol Eng 8(6):319–328. DOI: 10.1061/(ASCE)1084-0699(2003)8:6(319)

Kisi O, Uncuoglu E (2005) Comparison of three back-propagation training algorithms for two case studies. Indian J Eng Mater S 12:434-442

Kumar V, Panu U (1997) Predictive assessment of severity of agricultural droughts based on agro-climatic factors. J Am Water Resour Assoc 33(6):1255–1264. DOI: 10.1111/j.1752-1688.1997.tb03550.x

Levenberg K (1944) A method for the solution of certain non-linear problems in least squares. Q Appl Math 5:164–168

Liu WT, Negron-Juarez RI (2001) ENSO drought onset prediction in northeast Brazil using NDVI. Int J Remote Sens 22:3483–3501. DOI: 10.1080/01431160010006430

Lloyd-Hughes B, Saunders MA (2002) A drought climatology for Europe. Int J Climatol 22: 1571–1592. DOI: 10.1002/joc.846

Lohani VK, Loganathan GV (1997) An early warning system for drought management using the palmer drought index. J Am Water Resour Assoc 33(6):1375–1386. DOI: 10.1111/j.1752-1688.1997.tb03560.x

Lohani VK, Loganathan GV, Mostaghimi S (1998) Long-term analysis and short term forecasting of dry spells by the Palmer Drought Severity Index. Nord. Hydrol 29(1): 21–40

Mahmoudian, Z., and Baharvand, S. (2019). Investigating the Flow Pattern in Baffle Fishway Denil Type. Irrigation Sciences and Engineering, 42(3), 179–196. DOI: 10.22055/jise.2019.23693.1689.

Maier HR, Dandy GC (2000) Neural networks for the prediction and forecasting of water resources variables: a review of modelling issues and applications. Environ Model Softw. 15 (1):101-124. DOI: 10.1016/S1364-8152(99)00007-9

Marquardt D (1963) An algorithm for least-squares estimation of nonlinear parameters. SIAM J Appl Math 11(2):431–441

Martinez, J., Deng, Z. D., Tian, C., Mueller, R., Phonekhampheng, O., Singhanouvong, D., Thorncraft, G., Phommavong, T., and Phommachan, K. (2019). In situ characterization of turbine hydraulic environment to support development of fish-friendly hydropower guidelines in the lower Mekong River region. Ecological Engineering, Elsevier, 133(December 2017), 88–97.

McKee TB, Doesken NJ, Kleist J (1993) The relationship of drought frequency and duration to time scales. In proceedings of the 8th conference of Applied Climatology, 17-22 January 1993, American Meteorological Society, Anaheim, CA, pp 179–184

Mishra, AK, Desai VR (2005a) Drought forecasting using stochastic models. J Stoch Environ Res Risk Assess 19:326–339. DOI: 10.1007/s00477-005-0238-4

Mishra, AK, Desai VR (2006) Drought forecasting using feed forward recursive neural network. Ecol Model 198(1-2):127–138. DOI: 10.1016/j.ecolmodel.2006.04.017

Mishra AK, Desai VR, Singh VP (2007) Drought forecasting using a hybrid stochastic and neural network model. J Hydrol Eng 12 (6):626–638. DOI: 10.1061/ASCE1084-0699200712:6626

Mishra AK, Singh VP (2010) A review of drought concepts. J Hydrol 391(1–2): 202–216. DOI: 10.1016/j.jhydrol.2010.07.012

Moreira, EE, Paulo AA, Pereira LS, Mexia JT (2006) Analysis of SPI drought class transitions using loglinear models. J Hydrol 331:349–359. DOI: 10.1016/j.jhydrol.2006.05.022

Morid S, Smakhtin V, Moghaddasi M (2006) Comparison of seven meteorological indices for drought monitoring in Iran. Int J Climatol 26(7):971–985. DOI: 10.1002/joc.1264

Morid S, Smakhtin V, Bagherzadeh K (2007) Drought forecasting using artificial neural networks and time series of drought indices. Int J Climatol 27(15):2103–2111. DOI: 10.1002/joc.1498

Murphy BF, Timbal B (2007) A review of recent climate variability and climate change in Southeastern Australia. Int J Climatol 28 (7):859–879. DOI: 10.1002/joc.1627

Office for the Coordination of Humanitarian Affairs (OCHA) (2011) Eastern Africa Drought Humanitarian Report No. 3. United Nations Office for the Coordination of Humanitarian Affairs

Ozger M, Mishra AK, Singh VP (2012) Long lead time drought forecasting using a wavelet and fuzzy logic combination model: a case study in Texas. J Hydrometeorol 13(1):284–297. DOI: 10.1175/JHM-D-10-05007.1

OFDA/CRED (2008) EM-DAT: emergency events database. Univ. Catholique Louvain OFDA/CRED. Available at http://www.emdat.be/

Paulo AA, Ferreira E, Coelho C, Pereira LS (2005) Drought class transition analysis through Markov and Loglinear models, an approach to early warning. Agric Water Manage 77(1-3):59–81. DOI: 10.1016/j.agwat.2004.09.039

Paulo AA, Pereira LS (2007) Prediction of SPI drought class transitions using Markov chains. Water Resour Manag 21(10):1813–1827. DOI: 10.1007/s11269-006-9129-9

Rao AR, Padmanabhan G (1984) Analysis and modelling of Palmer drought index series. J Hydrol 68(1-4):211–229. DOI: 10.1016/0022-1694(84)90212-9

Raziei T, Saghafian B, Paulo AA, Pereira LS, Bordi I (2009) Spatial patterns and temporal variability of drought in western Iran. Water Resour Manag 23(3):439–455. DOI: 10.1007/s11269-008-9282-4

Raziei T, Bordi I, Pereira LS (2011) An application of GPCC and NCEP/NCAR datasets for drought variability analysis in Iran. Water Resour Manag 25(4):1075–1086. DOI: 10.1007/s11269-010-9657-1

Riebsame WE, Changnon SA, Karl TR (1990) Drought and natural resource management in the United States: Impacts and Implications of the 1987–1989 Drought. Westview, p. 174

Ross T, Lott N (2003) A Climatology of 1980–2003 extreme weather and climate events. National Climatic Data Center Technical Report No. 2003-01. NOAA/NESDIS. National Climatic Data Center, Asheville

Sen Z (1990) Critical drought analysis by second order Markov chain. J Hydrol 120(1-4):183–202. DOI: 10.1016/0022-1694(90)90149-R

Singh, A., Reager, J., and Behrangi, A. (2019). Estimation of hydrological drought recovery based on GRACE water storage deficit. Hydrology and Earth System Sciences Discussions, 40(December), 1–23.

Shabani, S., Yousefi, P., Naser, G. (2017). Support Vector Machines in Urban Water Demand Forecasting Using Phase Space Reconstruction, Procedia Engineering, 186: 537-543. https://doi.org/10.1016/j.proeng.2017.03.267.

Shiau JT, Modarres R (2009) Copula–based drought severity–duration–frequency analysis in Iran. Meteorol Appl 16(4):481–489. DOI: 10.1002/met.145

Steinemann A (2003) Drought indicators and triggers: a stochastic approach to evaluation. J Am Water Res Assoc 39(5):1217–1233. DOI: 10.1111/j.1752-1688.2003.tb03704.x

Sun Y, Wendi D, Kim DE, Liong S, (2016) Technical note: Application of artificial neural networks in groundwater table forecasting – a case study in a Singapore swamp forest. Hydrol. Earth Syst. Sci (20):1405-1412. DOI: 10.5194/hess-20-1405-2016

Tabari H, Abghari H, Hosseinzadeh Talaee P (2012) Temporal trends and spatial characteristics of drought and rainfall in arid and semiarid regions of Iran. Hydrol Process 26(22):3351–3361. DOI: 10.1002/hyp.8460

Thom HCS (1958) A note on gamma distribution. Mon Weather Rev 86(4):117–122

USAID/FEWSN (2011) Famine Early Warning System Network. East Africa: Past Year one of the Driest on Record in the Eastern Horn

Vijayalaksmi, D. P., & Babu, K. S. J. (2015). Water Supply System Demand Forecasting Using Adaptive Neuro-fuzzy Inference System. Aquatic Procedia. https://doi.org/10.1016/j.aqpro.2015.02.119

Wang W-C, Chau K-W, Cheng C-T, Qiu L (2009) A comparison of performance of several artificial intelligence methods for forecasting monthly discharge time series. J Hydrol 374:294–306. DOI: 10.1016/j.jhydrol.2009.06.019

Wilhite DA, Hayes MJ (1998) Drought planning in the United States: status and future directions. In: Bruins HJ, Lithwick H (eds) The Arid Frontier. Kluwer, Dordrecht

Wilhite DA, Hayes MJ, Svodoba MD (2000) Drought monitoring and assessment in the U.S. In: Voght JV, Somma F (eds) Drought and drought mitigation in Europe. Kluwers, Dordrecht

World Meteorological Organization (WMO) (1986) Report on drought and countries affected by drought during 1974–1985, Geneva, p. 118

Zarghami M, Abdi A, Babaeian I, Hassanzadeh Y, Kanani R (2011) Impacts of climate change on runoffs in East Azerbaijan, Iran. Global Planet Change 78(3–4):137–46. DOI: 10.1016/j.gloplacha.2011.06.003